\documentclass[%
reprint,
nofootinbib,  
 amsmath,amssymb,
 aps,
 pra,
]{revtex4-2}

\usepackage{graphicx}
\usepackage{dcolumn}
\usepackage{bm}
\usepackage{dsfont}  
\usepackage{bbold}  
\usepackage{empheq}  
\usepackage{empheq}  
\usepackage{amsmath}
\usepackage{ulem}
\usepackage[colorlinks=true,linkcolor=blue,citecolor=blue,urlcolor=blue]{hyperref} 

\DeclareMathOperator{\Tr}{Tr}
\DeclareMathOperator{\sinc}{sinc}
\DeclareMathOperator{\e}{e}
\newcommand{\ket}[1]{|#1\rangle}
\newcommand{\bra}[1]{\langle #1|}

\newcommand{\proj}[1]{|#1\rangle\langle #1|}

\newcommand{\modulus}[1]{|#1|}

\begin{document}

\preprint{APS/123-QED}

\title{Mode-selective single-photon addition to a multimode quantum field}

\author{Gana\"el Roeland$^1$}
\email{ganael.roeland@gmail.com}
\author{Srinivasan Kaali$^{1,2}$}
\author{Victor Roman Rodriguez$^3$}
\author{Nicolas Treps$^1$}
\author{Valentina Parigi$^1$}

\affiliation{$^1$Laboratoire Kastler Brossel, Sorbonne Universit\'e, ENS-Université PSL, CNRS, Coll\`ege de France, 4 place Jussieu, Paris F-75252, France}
\affiliation{$^2$ School of Quantum Technology, Defence Institute of Advanced Technology (DU), Girinagar, Pune, India}
\affiliation{$^3$Sorbonne Universit\'e, LIP6, CNRS, 4 place Jussieu, 75005 Paris, France}

\date{\today}

\begin{abstract}
Spectro-temporal modes of light can be exploited for the generation of high-dimensional Gaussian quantum states. Such states are at the basis of continuous variable quantum information protocols where they have to support mode-selective non-Gaussian operations. We develop a general framework for single-photon addition on multimode states of light via parametric down conversion processes. We identify the analytical conditions for single-mode and mode-selective photon addition. We show that spectral mode selectivity can be achieved in the Type-II collinear down conversion, while single-mode  condition are retrieved for noncollinear Type-I and Type-II processes. Numerical results are shown for photon addition in  parametric down conversion  process at near-infrared and telecommunications wavelengths. 
\end{abstract}

\maketitle

\tableofcontents


\section{Introduction}

Spectro-temporal mode of light are a versatile resource for quantum information and quantum communication protocols \cite{fabre,Brecht15,Karpinski21}. In particular ultra-fast light, that can be easily manipulated via femtosecond shaping techniques, has been used for application in both Discrete Variables (DV) and Continuous Variable (CV) encoding \cite{Zhang14,Roslund14,Kielpinski11,Lukens17,cai2017}
In order to exploit the large Hilbert space offered by the frequency mode of femtosecond light sources, the tailoring of the spectral mode structure for quantum state generation and manipulation should be performed. \cite{Roslund14, Ashby20,Brecht15,Brecht16,Patera12,ArzaniPRA,Roman-Rodriguez21,Eckstein11,ra2017,Allgaier20}.

In  CV quantum optics non-Gaussian quantum states are essential constituents for quantum computation \cite{Bart02,Mari12}. While spectrally tailored quantum states with Gaussian quadratures statistics can be deterministically generated via nonlinear optics \cite{Roslund14,Chen14,cai2017}, optical non-Gaussian states require heralded procedures like single-photon subtraction and single-photon addition. 

The two operations have been largely investigated acting on single-mode fields \cite{Lvovsky20}, where photon-subtraction can be implemented via a low-reflectivity beam-splitter \cite{Ourjoumtsev06}  and single-photon addition via a parametric amplifier with a strongly filtered heralding field \cite{Zavatta04}. 
 A general theoretical framework of mode selective single photon subtraction has been recently developed \cite{Averchenko14,Averchenko16} and experimentally demonstrated via sum-frequency conversion in nonlinear crystals \cite{ra2017,ra2020}. Differently from the low-reflectivity beam-splitter, the nonlinear frequency conversion allows for the subtraction of a single-photon from a selected ultra fast frequency mode of a multimode quantum state.  Single-photon addition has been recently implemented in delocalized temporal modes \cite{Biagi21} but spectral mode selectivity is still missing. 
 
The present work is focused on developing a complete theoretical framework to generate non-Gaussian quantum states of light by performing the addition of a single photon to multimode light fields.

We both analytically and numerically investigate under which conditions and experimental configurations it is possible to achieve single-mode and mode-selective photon addition, i.e. when one can arbitrarily choose the unique mode in which the photon is added.

We analyze configurations of parametric down-conversion (PDC) processes in nonlinear bulk crystals both at near infrared and telecommunication wavelength, which can be pumped via fields of different spectral shapes.



This paper is structured as follows. In section~\ref{sec:single_mode}, we briefly discuss photon addition to a single-mode light field. In section~\ref{sec:multimode}, we provide a complete theoretical description to the single-mode addition process in a mode selective way to a multimode light field, and discuss the output state purity. In section~\ref{sec:collinear_theory}, we analytically show that mode-selective photon addition is achievable in Type-II collinear PDC, and recover as necessary condition the group velocity matching (GVM) between the pump and one of the daughter field in the PDC process, which was already studied as beneficial condition for the generation of pure single-photons in spontaneous PDC \cite{mosley2008}. In section~\ref{sec:collinear_simu}, we show simulations in this configuration under realistic experimental conditions. In section~\ref{sec:noncollinear}, we study single-mode photon addition in noncollinear configurations by extending the group velocity matching condition to both Type-II (section~\ref{sec:noncollinear_typeII}) and Type-I (section~\ref{sec:noncollinear_typeI}). Though these results are valid for all classes of uniaxial and biaxial crystals, we present results for KDP, BBO, LN, BiBO and KTP crystals. Further discussions and prospects are given in conclusions.

\section{\label{sec:single_mode}Single-mode photon addition}

In this section, we will restrict ourselves to single-mode addition.
Single-photon added states have been introduced for the first time by~\cite{Agarwal92} and single photon addition has been implemented for the first time on a coherent state by~\cite{Zavatta04}.

\begin{figure}[h!]
	\begin{center}
		\includegraphics[width=0.8\columnwidth]{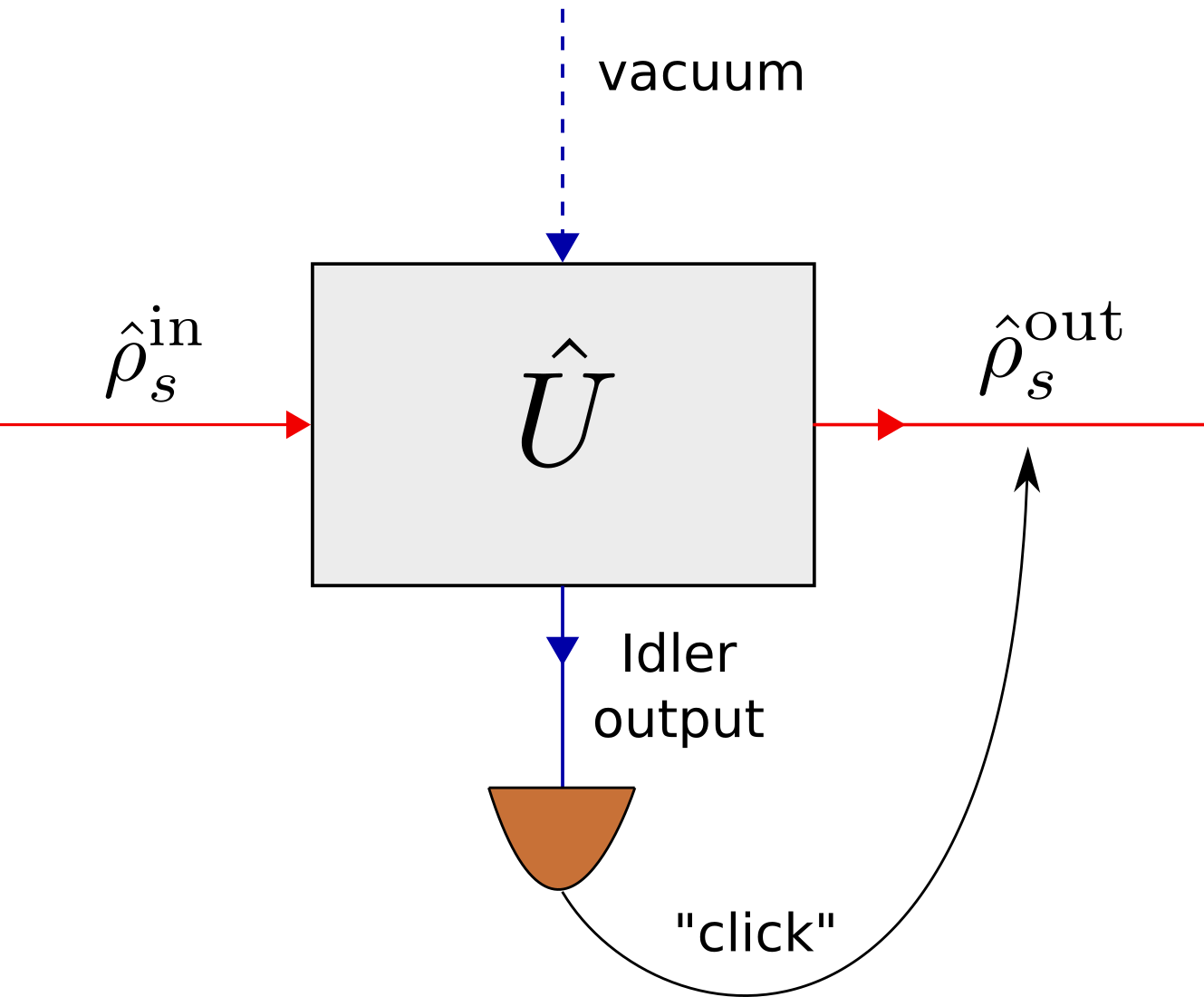}
	\end{center}
	\caption{\textbf{Principal scheme of conditional photon addition.}  The signal (resp. idler) channel is colored in red (resp. blue). The output state $\hat{\rho}^\mathrm{out}_s$ is conditioned on the detection of a photon in the idler channel.}\label{fig:addition_single_mode}
\end{figure}

We consider a parametric generation process, where we adopt a simplified model of parametric down conversion in a nonlinear crystal. The process is illustrated on Fig.~\ref{fig:addition_single_mode}. At the input, the quantum beam of light to which we want to add a photon, is called the signal. The associated quantum state is described by the general density matrix $\hat{\rho}^\mathrm{in}_s$. The process, modelled by its evolution operator $\hat{U}$, generates two photons, one in the signal channel and a complementary one in the channel called idler. 
The output $\hat{\rho}^\mathrm{out}_s$ on the signal channel is conditioned by the detection of a photon in the idler channel.

The evolution operator writes: 
\begin{equation}\label{eq:single-mode_addition_evol_op}
\hat{U}=\exp(g (\hat{a}\hat{b}-\hat{a}^\dag\hat{b}^\dag) )\approx \hat{\mathds{1}} + g (\hat{a}\hat{b}-\hat{a}^\dag\hat{b}^\dag)
\end{equation}
where g is the strength of the parametric generation, containing the non-depleted pump, $\hat{a}$ is the annihilation operator associated to the signal mode, and $\hat{b}$ is the annihilation operator associated to the idler mode. In equation~\eqref{eq:single-mode_addition_evol_op}, we have assumed that the coupling is weak, i.e. $\modulus{g} \ll 1$. By applying the evolution operator to the total input state $\hat{\rho}^\mathrm{in}~\!=~\!\hat{\rho}^\mathrm{in}_s~\!\otimes~\!\proj{0}_i$, we obtain the evolution equation:
\begin{equation}
\hat{U} \hat{\rho}^\mathrm{in} \hat{U}^\dag = \hat{\rho}^\mathrm{in} - \modulus{g}^2 \hat{a}^\dag \hat{b}^\dag \hat{\rho}^\mathrm{in} \hat{a} \hat{b}
\end{equation}
Since the possibility to add more than one photon to the signal is negligible in the weak coupling approximation, one can model the on/off detector as $\hat{\Pi} = \hat{\mathds{1}}_i - \proj{0}_i \approx \proj{1}_i \ $. The output density matrix in the signal mode, conditioned to the measurement of a photon in the idler mode is formally given by:
\begin{equation}
\hat{\rho}^\mathrm{out}_s = \frac{1}{P} \Tr_i (\hat{\Pi}\,\hat{U} \hat{\rho}^\mathrm{in}\, \hat{U}^\dag)
\end{equation}
where the normalization constant $P = \Tr_{i,s} (\hat{\Pi}\,\hat{U} \hat{\rho}^\mathrm{in}\, \hat{U}^\dag) = \modulus{g}^2 (1+\bar{n}_s)$ is the probability to successfully detect a photon in the idler mode \cite{Zavatta07}, with $\bar{n}_s = \Tr(\hat{a}^\dag \hat{a} \hat{\rho}^\mathrm{in}_s )$ the mean number of photons in the input state. We point out that the probability to detect a photon depends linearly on $1+\bar{n}_s$, which reflects that parametric down conversion behaves as an amplifier. Indeed, the more photons are in the input state, the more likely it is to add a photon to the signal.

Finally, we obtain that:
\begin{equation}\label{eq:single-mode_rho_out}
\hat{\rho}^\mathrm{out}_s = \frac{ \hat{a}^\dag \hat{\rho}^\mathrm{in}_s \hat{a} }{ 1+ \bar{n}_s }
\end{equation}
The output signal state in Eq.~\eqref{eq:single-mode_rho_out} is as expected: the input state on which is added a photon by applying $\hat{a}^\dag$.


\section{\label{sec:multimode}Multimode photon addition}

\subsection{\label{sec:multimode_frame}General framework}

In this section, we extend the simple previous theory to the multimode case: we consider that photons can be added to any mode. Fig.~\ref{fig:addition_single_mode} still describes the general setting of the process.

The evolution operator of multimode parametric generation in the low gain regime writes:
\begin{equation}\label{eq:op_ev}
\hat{U} \approx \hat{\mathds{1}} + \sum_{n,m} (g_{nm} \hat{a}^\dag_n \hat{b}^\dag_m + \mathrm{h.c.})
\end{equation}
where we define the signal modes as the optical modes $\{u_n\}$ associated to the annihilation operators $\{\hat{a}_n\}$, and the idler modes as the optical modes $\{v_n\}$ associated to the annihilation operators $\{\hat{b}_n\}$, $g_{nm}$ is the strength of the process for modes $(u_n,v_m)$, and $\mathrm{h.c.}$ stands for hermitian conjugate. Note that $\hat{U}$ can be derived from a Hamiltonian approach \cite{fabre,Par07}. Again, the input writes $\hat{\rho}^\mathrm{in}~\!=~\!\hat{\rho}^\mathrm{in}_s~\!\otimes~\!\proj{0}_i$, where $\hat{\rho}^\mathrm{in}_s$ is the signal input, potentially mixed or multimode.

Let's compute the output signal state conditioned on the detection of a photon in the idler beam. Here we assume to be in the weak coupling regime ($|g_{nm}| \ll 1, \ \forall n,m$ ) in order to neglect the possibility to add more than one photon. In this regime, we can model the on/off detector as:
\begin{equation}\label{eq:detection_simplified}
\hat{\Pi} = \hat{\mathds{1}}_i - \proj{0}_i \approx \sum_d \proj{1}_{r_d}
\end{equation}
where the detection mode $r_d$ are associated to the annihilation operator $\hat{D_d}$ and $\ket{1}_{r_d} = \hat{D_d}^\dag\ket{0}$.


The output state simply writes $\hat{\rho}^\mathrm{out} = \hat{U} \hat{\rho}^\mathrm{in}\, \hat{U}^\dag$.
The conditional output signal after a successful click on the detector is:
\begin{equation}
\hat{\rho}_s^\mathrm{out} = \frac{1}{P}\; \Tr_i (\hat{\Pi}\,\hat{U} \hat{\rho}^\mathrm{in}\, \hat{U}^\dag)
\end{equation}
where the normalization constant $P = \Tr_{i,s} (\hat{\Pi}\,\hat{U} \hat{\rho}^\mathrm{in}\, \hat{U}^\dag )$ is the probability to successfully detect a photon in the idler mode. 
One can show that the output signal state writes:
\begin{align}\label{eq:rho_out_s_general}
\hat{\rho}^\mathrm{out}_s &= \frac{1}{P}\; \sum_{n,n'} A_{nn'} \hat{a}^\dag_n \hat{\rho}^\mathrm{in}_s\, \hat{a}_{n'}\\
A_{nn'} &= \sum_m g_{nm}g_{n'm}^* \label{eq:definition_addition_matrix}
\end{align}

The behaviour of this whole process is governed by the matrix $(A_{nn'})$, which we will refer to as the addition matrix. Note that the addition matrix is hermitian by definition~\eqref{eq:definition_addition_matrix}. The diagonalization of the addition matrix gives access to the eigenvalues $\lambda_1 \geq \dots \geq \lambda_n \geq 0$ and the eigenmodes $\{w_n\}$ associated to the annihilation operators $\{\hat{e}_n\}$. We obtain:
\begin{align}
\hat{\rho}^\mathrm{out}_s &= \frac{1}{P}\sum_n \lambda_n \hat{e}^\dag_n \hat{\rho}^\mathrm{in}_s \, \hat{e}_n \label{eq:rho_out_s_diag}
\\ \mathrm{where}\quad P &= \sum_n \lambda_n ( 1 + \bar{n}_n) \label{eq:rho_out_s_diag_norm}
\end{align}
and $\bar{n}_n = \Tr (\hat{e}^\dag_n \hat{e}_n \hat{\rho}^\mathrm{in}_s )$ is the photon number of the input signal in the eigenmode $w_n$.

In the general case, the addition process is multimode, i.e. more than one eigenvalue $\lambda_n$ is non-zero. The effective number of modes in the process is given by the following quantity, which is similar to the Schmidt number~\cite{ekert1995}:
\begin{equation}\label{eq:schmidt_number}
 K = \frac{(\sum_n \lambda_n)^2}{\sum_n \lambda_n^2}
\end{equation}
The addition process is single-mode when $K=1$.

We will now discuss in detail those two cases, looking into their link with the output state purity. Intuitively, the purity of the output state decreases as the total number of modes involved in the process increases, since the single photon can be added into more eigenmodes, following Eq.~\eqref{eq:rho_out_s_diag}.

\subsection{\label{multi_add_purity} Output state purity}

In this section, we assume that the input signal is pure, i.e. $\hat{\rho}^\mathrm{in}_s = \proj{\phi}_s$.

First, if the addition process is single-mode, then Eq.~\eqref{eq:rho_out_s_diag} simply re-writes:
\begin{equation}
\hat{\rho}^\mathrm{out}_s \propto \hat{e}^\dag_1 \hat{\rho}^\mathrm{in}_s \, \hat{e}_1 = \hat{e}^\dag_1 \proj{\phi}_s \hat{e}_1
\end{equation}
We deduce that the output signal is pure: $\hat{\rho}^\mathrm{out}_s = \proj{\psi}_s$, with $\ket{\psi} \propto \hat{e}^\dag_1 \ket{\phi}$. A photon has been properly added to the eigenmode. This is the ideal single-mode photon addition process, at $K=1$.

Let's now consider the case of a multimode process, i.e. $K>1$. We will show that the output is always mixed once the addition process is multimode.
\\For simplicity, we first assume that only two eigenvalues are non-zero. 
Then, equations~\eqref{eq:rho_out_s_diag}~and~\eqref{eq:rho_out_s_diag_norm} rewrite:
\begin{align}
&\hat{\rho}^\mathrm{out}_s = \tilde{\lambda}_1 \hat{e}_1^\dag \hat{\rho}^\mathrm{in}_s \hat{e}_1 + \tilde{\lambda}_2 \hat{e}_2^\dag \hat{\rho}^\mathrm{in}_s \hat{e}_2 \label{eq:2_eigenvalues_rho}\\
&\tilde{\lambda}_1 (1+\bar{n}_1) + \tilde{\lambda}_2 (1+\bar{n}_2) = 1 \label{eq:norm}\\
\mathrm{where}\quad &\tilde{\lambda}_i = \lambda_i/P\quad\mathrm{for}\  i=1,2 \nonumber
\end{align}
As $\hat{\rho}^\mathrm{in}_s$ is pure, we find, using trace properties, that the output state purity writes:
\begin{equation}\label{eq:2_eigenvalues_purity}
\Tr [ (\hat{\rho}^\mathrm{out}_s)^2 ] = \tilde{\lambda}_1^2 (1+\bar{n}_1)^2 + \tilde{\lambda}_2^2 (1+\bar{n}_2)^2 + 2 \tilde{\lambda}_1 \tilde{\lambda}_2 \left| \bra{\phi} \hat{e}_1 \hat{e}_2^\dag \ket{\phi} \right|^2
\end{equation}
We apply the Cauchy-Schwarz inequality:
\begin{equation}\label{eq:cauchy_schwarz}
|\bra{\phi} \hat{e}_1 \hat{e}_2^\dag \ket{\phi}|^2 \leq \bra{\phi} \hat{e}_1 \hat{e}^\dag_1 \ket{\phi} \bra{\phi} \hat{e}_2 \hat{e}^\dag_2 \ket{\phi} = (1+\bar{n}_1)(1+\bar{n}_2)
\end{equation}
This allows us to write:
\begin{equation}\label{eq:CS}
\Tr [ (\hat{\rho}^\mathrm{out}_s)^2 ] \leq (\tilde{\lambda}_1 (1+\bar{n}_1) + \tilde{\lambda}_2 (1+\bar{n}_2))^2 = 1
\end{equation}
where we used the normalisation equation~\eqref{eq:norm}. The inequality~\eqref{eq:CS} is saturated if and only if $\hat{e}^\dag_1 \ket{\phi} \propto \hat{e}^\dag_2 \ket{\phi}$, which is not possible (see proof in Appendix~\ref{app:A_cauchy}). So the purity of the output signal density matrix is strictly lower than $1$, meaning that the output signal state is mixed.
\\This result can be generalized to more than two non-zero eigenvalues, without any additional steps, as can be checked on Appendix~\ref{app:A_genral_purity}.
\\Therefore, we have shown that for a \textbf{multimode addition process} ($K \neq 1$), for \textbf{any} input signal \textbf{the heralded output signal is not pure}.

\begin{figure}[t!]  
	\begin{center}
		\includegraphics[width=1\columnwidth]{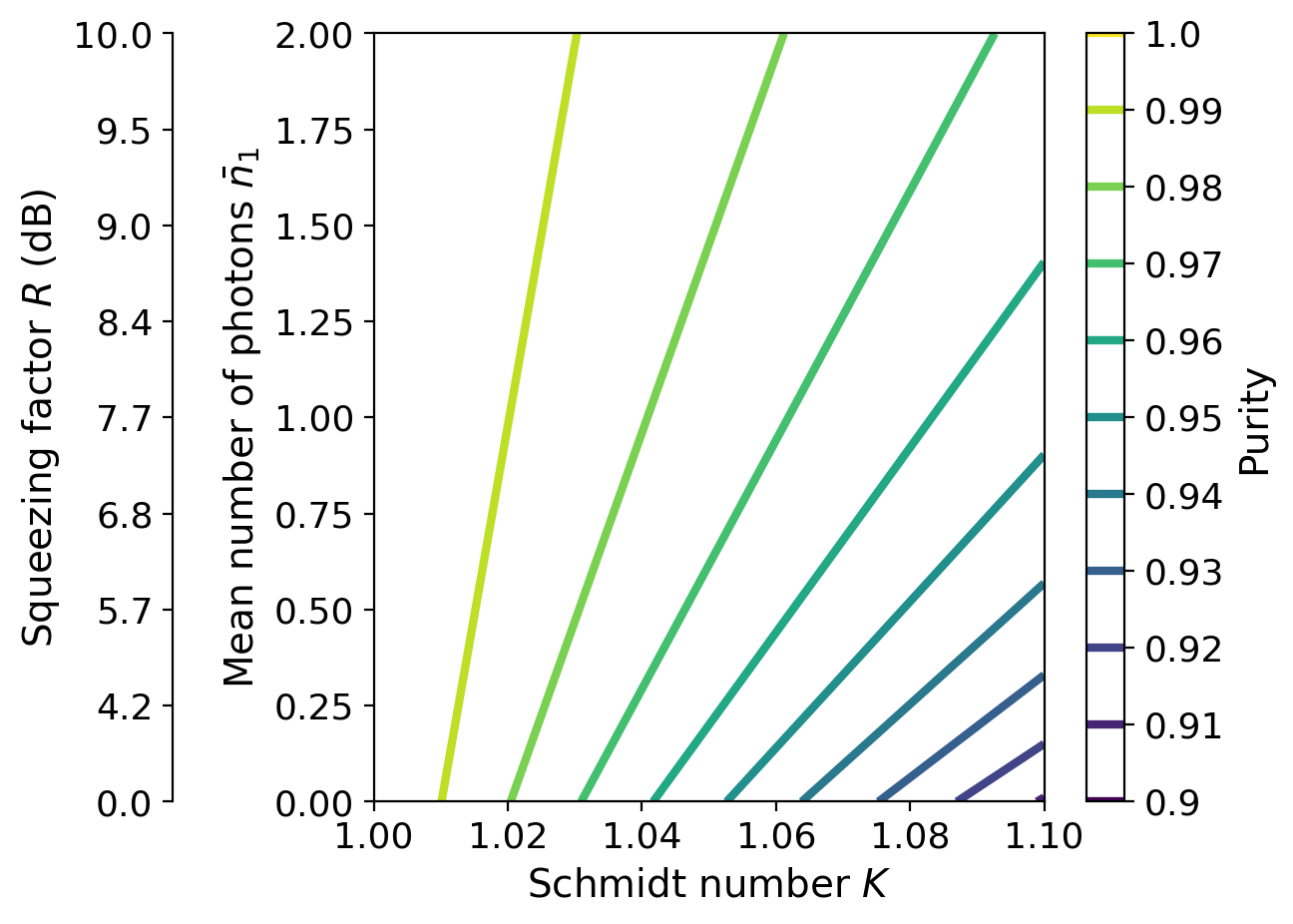}
	\end{center}
	\caption{\textbf{Output state purity for a single-mode input state.} At $K=1$, the process is single-mode, which leads to a purity equal to $1$. Below, the purity drops with $K$, and increases with $\bar{n}_1$. The squeezing factor axis is non linear, and is derived from $\bar{n}_1 = \text{sinh}^2(R)$.}\label{fig:purity_sensibility1}
\end{figure}

Now, let's quantitatively study the dependence of the output state purity on some relevant input states. In quantum information experiments, state purity is a very sensitive parameter and need to be as close to $1$ as possible.  
\\In this section, to simplify the discussion, we consider that the process is mainly determined by two eigenmodes, so that the output state is given by Eq.~\eqref{eq:2_eigenvalues_rho}.  

We consider the situation where the input signal is a pure single-mode state:
\begin{equation}\label{purity_sensibility_state}
\hat{\rho}^\mathrm{in}_s = \proj{\phi}, \quad \ket{\phi} = \ket{\chi}_1 \ket{0}_2
\end{equation}
where in the eigenmode $w_1$, $\ket{\chi}_1$ has $\bar{n}_1$ mean number of photons.
Using that the scalar product $\bra{\phi} \hat{e}_1 \hat{e}_2^\dag \ket{\phi}$ vanishes and $\bar{n}_2=0$ in equations~\eqref{eq:norm}~and~\eqref{eq:2_eigenvalues_purity}, we obtain:
\begin{equation}\label{eq:output_purity_mono_squeezed}
\Tr [ (\hat{\rho}^\mathrm{out}_s)^2 ] = \frac{1 +  (\tilde{\lambda}_1/\tilde{\lambda}_2)^2(1+\bar{n}_1)^2 }{[1 + (\tilde{\lambda}_1/\tilde{\lambda}_2) (1+ \bar{n}_1)]^2}
\end{equation}

We show in Fig.~\ref{fig:purity_sensibility1} the output state purity as a function of the effective number of modes $K=(1+\tilde{\lambda}_1/\tilde{\lambda}_2)^2/(1+(\tilde{\lambda}_1/\tilde{\lambda}_2)^2)$, and mean number of photons $\bar{n}_1$. The figure illustrates the competition between the multimodality and the amplification effect of the process. We point out that for an effective number of modes smaller or equal to 1.1, the purity is always above about 0.90. Note that Eq.~\eqref{eq:output_purity_mono_squeezed} is true for any state of the modal form Eq~\eqref{purity_sensibility_state}. The usual candidate states for photon addition are coherent, thermal and squeezed states. Squeezed states are of particular interest for quantum information, as they can be entangled into a cluster, building block of measurement based quantum computing. Adding a photon to a squeezed state leads to a non-Gaussian resource, necessary for quantum computation \cite{mari2012}.  We show on a secondary axis the squeezing factor $R$, related to the mean number of photons by $\bar{n}_1 = \text{sinh}^2(R)$ for single-mode squeezed vacuum states. 

The fact that a non-pure state can emerge from photon addition on a pure single-mode state essentially comes from the non-zero probability of adding a photon to the vacuum. In comparison, in the photon subtraction process \cite{Averchenko16}, the output is always pure if the input signal is pure and single-mode, as subtracting from the vacuum is impossible.

\section{\label{sec:collinear}Mode-selective photon addition in Type-II collinear PDC}

In this section, we develop an experimental model of the addition matrix $A_{n,n'}$, and diagonalize it both analytically and numerically. The goal is to find:
\begin{itemize}
\item under which conditions the process can be single-mode, meaning that the effective number of modes in which it adds a photon is reduced to one.
\item  under which conditions the process can be mode-selective, meaning that one can choose in which eigenmode the photon is added.
\end{itemize}

\begin{figure}[t!]
	\begin{center}
		\includegraphics[width=1\columnwidth]{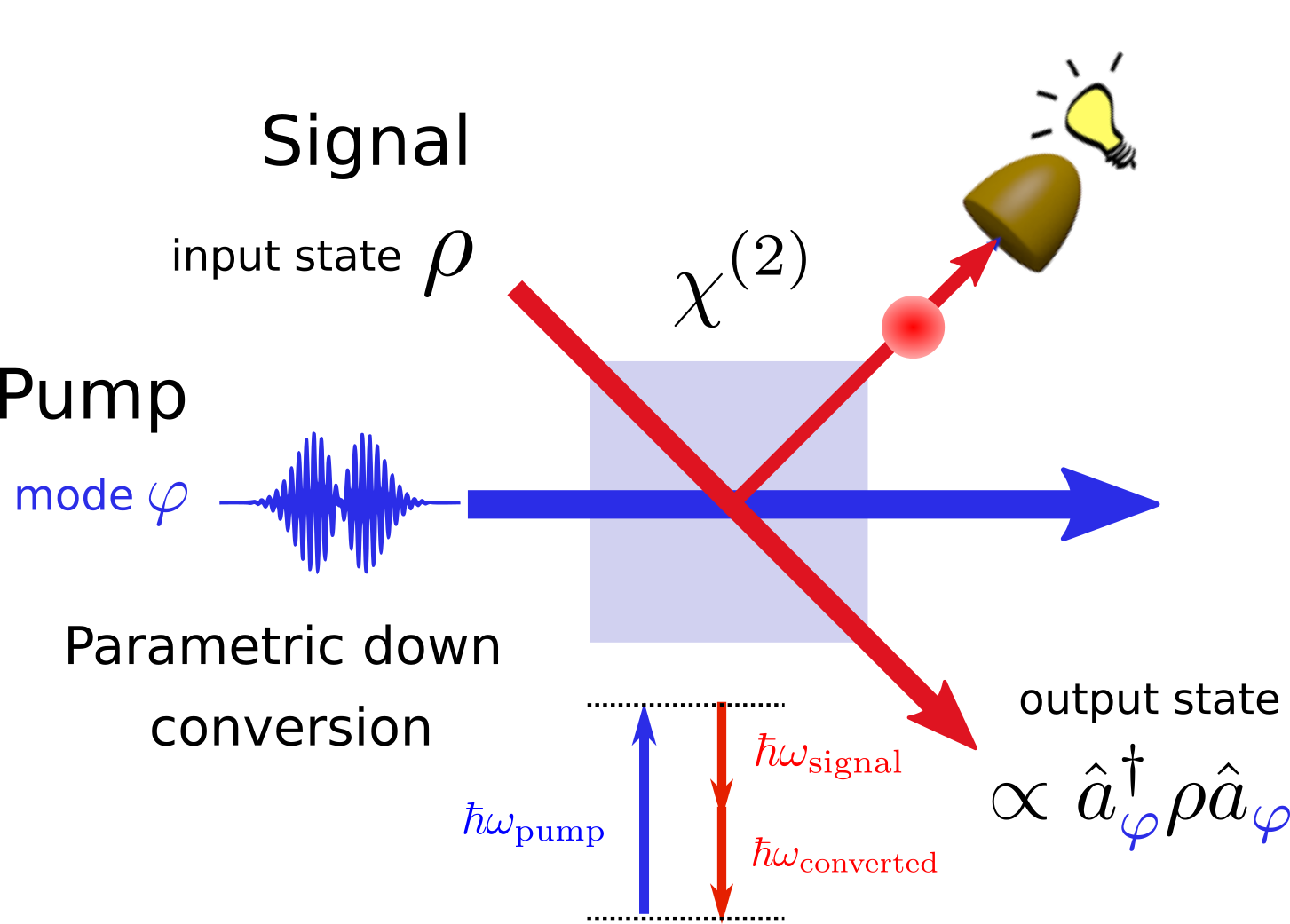}
	\end{center}
	\caption{\textbf{Mode-selective photon addition through parametric down conversion in a nonlinear crystal.} Each pump photon is down converted into one photon added to the signal and one idler photon detected for heralding purposes. Photon addition occurs in the spectral mode $\varphi$ of the pump.}\label{fig:exp_add}
\end{figure}

While the general principle of the process remains the same as described in Fig.~\ref{fig:addition_single_mode}, we now consider usual physical systems: parametric down conversion (PDC) in a nonlinear crystal, using pulsed light for both signal and pump beams. The modes at play are frequency modes of the usually large spectrum pulses. The pump is a classical beam that feeds the nonlinear crystal at the input, and is part of the process described by $\hat{U}$.
\\In this section, we focus on collinear type-II parametric down conversion, see Fig.~\ref{fig:exp_add}. Collinear means that input and output fields are all propagating in the same direction (on Fig.~\ref{fig:exp_add} they are not for clarity purposes). Being type-II means that the signal and idler output fields have orthogonal polarisations. This allows separating the output beams in practice. In sections \ref{sec:noncollinear_typeII} and \ref{sec:noncollinear_typeI}, we investigate noncollinear parametric down conversion configurations.

\subsection{\label{sec:collinear_theory}Mode-selectivity}
The evolution operator for a 3 wave mixing process in a nonlinear optical crystal, under the low gain approximation, is a continuous version of that of Eq.\eqref{eq:op_ev} \cite{Brecht15,ansari2018}:
\begin{align}
&\hat{U} \approx \hat{\mathds{1}} +  \int d\omega_sd\omega_i \left( R(\omega_s, \omega_i) \hat{a}^\dag(\omega_s) \hat{b}^\dag(\omega_i)  + \mathrm{h.c.} \right)\\
&R(\omega_s, \omega_i)  = (- 2 i \pi \mathcal{C}/ \hbar) \: \alpha_p(\omega_s+\omega_i) \, \phi(\omega_s, \omega_i)\label{eq:JSA_function_definition}
\end{align}
where  $R(\omega_s, \omega_i)$ is called the joint spectral amplitude (JSA) function, and $\phi(\omega_s, \omega_i) = \sinc\left( \Delta k L / 2 \right)$ is the phasematching function with the crystal length $L$, $\Delta k = k_p - k_s - k_i$, the frequencies $\omega_j$ and the wave vectors $k_j$ of the fields for $j=p,s,i$, and $\mathcal{C}$ is a constant\footnote{$\mathcal{C} = L \sqrt{\frac{W_p \hbar \omega_s^0 \hbar \omega_i^0}{8\epsilon_0^3 n_p n_s n_i c^3}} $, where $W_p$ is the energy contained in a single pulse of the field, $n_j$ (resp. $\omega_j^0$) is the refractive index seen by the fields (resp. the central frequencies) for $j=p,s,i$, $\epsilon_0$ is the vacuum permittivity, and $c$ is the speed of light.}. Since the evolution operator has a similar form as in Eq.~\eqref{eq:op_ev}, we find:
\begin{align}
\hat{\rho}^\mathrm{out}_s &= \frac{1}{P}\; \int d\omega_sd\omega_s' A(\omega_s, \omega_s') \hat{a}^\dag(\omega_s) \hat{\rho}^\mathrm{in}_s\, \hat{a}(\omega_s') \\
A(\omega_s, \omega_s') &= \int d\omega_i R(\omega_s, \omega_i) R(\omega_s', \omega_i)^* \label{eq:JSI_JSA_relation}
\end{align}
where we recall the input state form $\hat{\rho}^\mathrm{in}~\!=~\!\hat{\rho}^\mathrm{in}_s~\!\otimes~\!\proj{0}_i$ and that $P$ is a normalisation factor that ensures $ \Tr \left( \hat{\rho}^\mathrm{out}_s \right)=1$. Again, we can diagonalize $A(\omega_s, \omega_s')$, as it is hermitian and obtain the exact same equation as Eq.~\eqref{eq:rho_out_s_diag}, by finding the eigenmodes and eigenvalues :
\begin{empheq}[left={\empheqlbrace}]{equation} \label{eq:addition_matrix_diagonalized}
\begin{aligned}
\: &A(\omega_s, \omega_s') = \sum_{n\geq 1} \lambda_n \varphi_n(\omega_s) \varphi_n^*(\omega_s') \\
\: &\hat{e}^\dag_n = \int d\omega_s \varphi_n(\omega_s) \hat{a}^\dag(\omega_s)
\end{aligned}
\end{empheq}
where $\{\varphi_n(\omega_s)\}$ are the signal frequency eigenmodes. 
\\Such full diagonalization seems to be out of range of analytical computation. Yet, we show that, under some approximations, one can compute analytically an estimation of the effective number of modes $K$, defined in Eq.~\eqref{eq:schmidt_number}. We first make a gaussian approximation on the phasematching function, and assume a gaussian pump spectrum:
\begin{align}\label{eq:gaussian_phasematching}
\phi(\omega_s, \omega_i) &\approx \exp\left( - \gamma \left(\Delta k (\omega_s, \omega_i)L/2\right)^2\right) \\
\alpha_p(\omega_s + \omega_i) &\propto \exp\left(-\frac{(\tilde{\omega}_s+\tilde{\omega}_i)^2}{2\sigma^2}\right)  \label{eq:gaussian_pump}
\end{align}
where $\gamma \simeq 0.193$ is defined such that the functions $\sinc (x)$ and $\e^{-\gamma x^2}$ have the same full width at half maximum (FWHM), $\sigma$ is the pump spectral width, and $\tilde{\omega}_j = \omega_j - \omega_j^0$ for $j = p, s, i$ are the frequency shifts with $\omega_j^0$ the central frequency of each pulsed beam.
\\If we make the Taylor expansion of the phase mismatch $\Delta k (\omega_s, \omega_i)$ around the central frequencies , and we keep  up to the first order in $\tilde{\omega}_j$, we can write:
\begin{equation} \label{eq:delat_k_taylor}
\Delta k (\omega_s, \omega_i) = (k_p' - k_s') \tilde{\omega}_s + (k_p' - k_i') \tilde{\omega}_i + O(\tilde{\omega}^2)
\end{equation}
where $k_j' \equiv \frac{\partial k_j}{\partial \omega_j}|_{\omega_j^0}$ are the inverse of the field group velocities and where we have assumed perfect phasematching at the central frequencies, ($\Delta k (\omega_s^0, \omega_i^0)~=~0$). 
\\Substituting equations~\eqref{eq:gaussian_phasematching},~\eqref{eq:gaussian_pump}~and~\eqref{eq:delat_k_taylor} into the JSA function~\eqref{eq:JSA_function_definition} leads to:
\begin{align}\label{eq:JSA_gaussian_form}
R(\omega_s, \omega_i)  \propto \exp\Bigl(& - \frac{(\tilde{\omega}_s+\tilde{\omega}_i)^2}{2\sigma^2} \nonumber
\\ & - \frac{\gamma L^2}{4} \left((k_p' - k_s') \tilde{\omega}_s + (k_p' - k_i') \tilde{\omega}_i \right)^2\Bigr)
\end{align}
We show that under gaussian approximations, $K$ has an explicit analytical form (see Appendix \ref{app:B_derivation_K} for a detailed proof):
\begin{align}
K &= \sqrt{\frac{(1+r_s^2)(1+r_i^2)}{(r_s-r_i)^2}} \label{eq:efficient_nb_mode_computed} \\
\mathrm{with}\quad r_j &= \sigma L \sqrt{\frac{\gamma}{2}} \, | k_p' - k_j'|,\quad \mathrm{for} \ j = i, s \nonumber
\end{align}
where the adimensional $r_j$ coefficients contains all the key parameters of the problem. Expression~\eqref{eq:efficient_nb_mode_computed} allows quantifying of the multimodality of the addition process. In particular, this analytical computation allows us to find under which conditions the process is single-mode, i.e. $K=1$. Indeed to obtain $K=1$ from equation~\eqref{eq:efficient_nb_mode_computed}, one of the $r_j$ must vanish. Since we are interested into the selectivity over the signal mode, we choose $r_s=0$ similarly to \cite{mosley2008}. This leads to the group velocity matching condition:
\begin{equation}
k_p' = k_s'\quad \mathrm{(group\ velocity\ matching)} \label{eq:group_velocity_matching_condition}
\end{equation}
This condition can be achieved in some usual crystals, which is discussed in the next section~\ref{sec:collinear_simu}. $K$ now rewrites into:
\begin{equation}
K = \sqrt{1+\frac{1}{r_i^2}} \approx 1 \qquad \mathrm{if}\ r_i^2 \gg 1
\end{equation}
The single-mode condition $r_i^2 \gg 1$ leads to:
\begin{equation}\label{eq:condition_width}
\sigma^2 \gg \frac{1}{\gamma L^2 (k_p' - k_i')^2 /2}
\end{equation}
Note that condition~\eqref{eq:condition_width} can be physically seen as a long enough crystal condition or equivalently as a broad enough pump spectrum. The addition process is single-mode under conditions~\eqref{eq:condition_width}~and~\eqref{eq:group_velocity_matching_condition}.

For the process to be mode-selective, the output signal mode should be controllable by an experimental parameter: here it is the pump spectrum. Let's have a general not anymore gaussian pump spectrum $\alpha_p$, and show that we still have a single-mode process under conditions~\eqref{eq:group_velocity_matching_condition} and~\eqref{eq:condition_width}. Under the GVM condition, we can rewrite Eq.~\eqref{eq:JSA_gaussian_form} as:
\begin{equation}\label{eq:JSA_gaussian_PM}
R(\omega_s, \omega_i) \propto \alpha_p(\omega_s + \omega_i) \exp\left( - \frac{\gamma L^2}{4} (k_p' - k_i')^2 \tilde{\omega}_i^2\right)
\end{equation}
Under condition~\eqref{eq:condition_width}, the pump varies slowly in comparison to the phasematching function with respect to the variable $\omega_i$: $\alpha_p(\omega_s + \omega_i) \approx \alpha_p(\omega_s+\omega_i^0)$. Now, the JSA function can be written in a factorized from:
\begin{equation}\label{eq:R_factorized}
R(\omega_s, \omega_i)  \propto \alpha_p(\omega_s+\omega_i^0) \exp\left( - \frac{\gamma L^2}{4} (k_p' - k_i')^2 \tilde{\omega}_i^2\right)
\end{equation}
Let's compare it to the Schmidt decomposition of the JSA function into the signal and idler frequency eignemodes:
\begin{equation}\label{eq:JSA_function_Schmidt_decomposition}
R(\omega_s, \omega_i) = \sum_{n \geq 1} \sqrt{\lambda_n} \varphi_n(\omega_s) \psi_n^*(\omega_i)
\end{equation}
It is clear that in Eq.~\eqref{eq:R_factorized} the JSA function is reduced to a product of the form $R(\omega_s, \omega_i) \propto \varphi(\omega_s)\psi^*(\omega_i)$, with $\varphi$ (resp. $\psi$) the unique signal (resp. idler) eigenmode.
The addition process is thus single-mode and the signal mode $\varphi$ is given by the spectral shape $\alpha_p$ of the pump, i.e. $\varphi = \alpha_p$. In other words, the photon is added to the mode of the signal that has the same spectral shape as the pump. The mode of the pump can then be tailored via ultrafast shaping in order to choose the addition mode for the signal.


We conclude from these analytical considerations that the collinear PDC addition process is single-mode and mode-selective under group velocity matching condition~\eqref{eq:group_velocity_matching_condition} and broad enough pump spectrum or equivalently long enough crystal condition~\eqref{eq:condition_width}.

\subsection{\label{sec:collinear_simu}Simulations}
    
    This section is dedicated to finding the single-mode addition conditions for collinear Type-II PDC in nonlinear bulk crystals, with realistic parameters. In the next section we will extend it to the noncollinear case.
    
     
In Type-II PDC, for uniaxial bulk crystals, phasematching at signal and idler's central frequencies can be achieved with:
\begin{equation}\label{eq:T2C-PM-condition}
    2 n_{e}(\lambda_p, \theta_c) = n_{e}(\lambda_i,\theta_c) + n_{o}(\lambda_s)
\end{equation}
where $\theta_c$ is the crystal cut angle defined as the angle between the pump and the optical axis. At given wavelengths, this condition is satisfied if the crystal is cut at a specific angle called the phase matching angle, $\theta_c = \theta_{PM}$.

\begin{figure}[t!]
	\begin{center}
		\includegraphics[width=1\columnwidth]{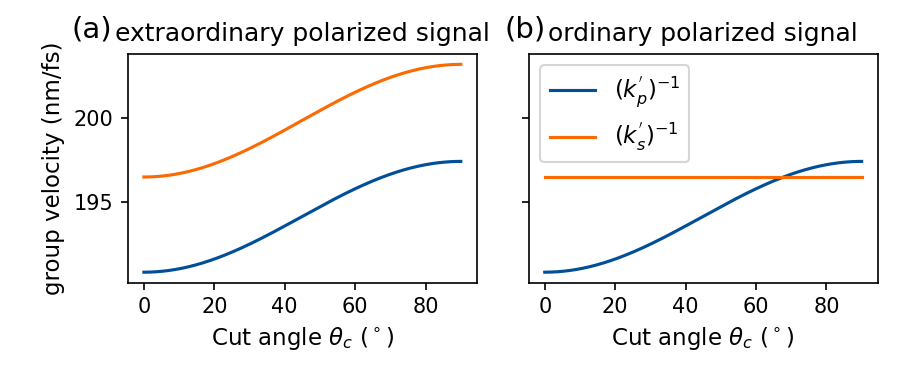}
	\end{center}
	\caption{Group velocity curves of the pump and signal fields for degenerate Type-II PDC in KDP. The pump is extraordinary polarized in both graphs. The signal is (a) extraordinary polarized, or (b) ordinary polarized.}
	\label{fig:GVMcurve}
\end{figure}

\begin{figure}[t!]
	\begin{center}
		\includegraphics[width=1\columnwidth]{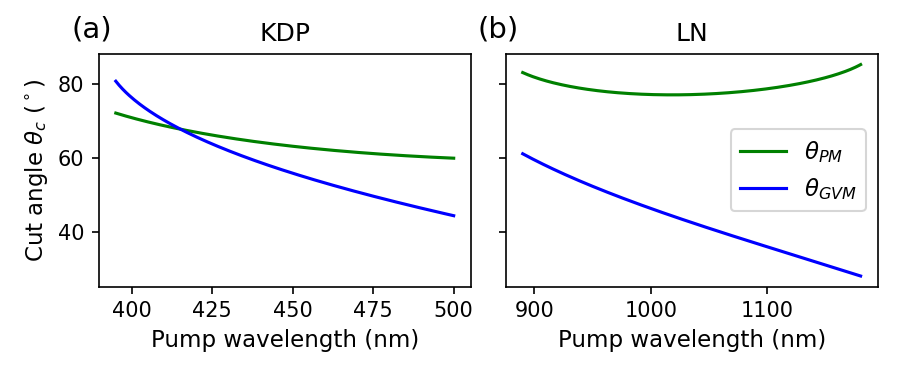}
	\end{center}
	\caption{Group velocity matching and phase matching curves for different pump wavelengths, for (a) KDP and (b) LN crystals. For LN, there is no solution to Eqs.~\eqref{GVM=PM}.}
	\label{fig:KDP-LN}
\end{figure}

As discussed earlier, the GVM condition is satisfied if $k'_{p} = k'_{s}$. For uniaxial crystals, the signal photon can be chosen as ordinary or extraordinary polarized. 
for KDP crystal, it is not possible to achieve the GVM condition for an extraordinary polarized signal field, see Fig.~\ref{fig:GVMcurve}. When the signal photon is ordinary polarized, however, the group velocities of the pump and the signal matches for a particular cut angle $\theta_c = \theta_{GVM}$.

For a given wavelength $\lambda_p$ of the pump photon, to achieve both the phasematching condition and the GVM condition, it requires that:
\begin{empheq}[left={\empheqlbrace\: }]{equation}
    \begin{aligned}
        \theta_c &= \theta_{PM}\\
        \theta_c &= \theta_{GVM}
        \label{GVM=PM}
    \end{aligned}
\end{empheq}
This condition cannot be achieved for an arbitrary pump wavelength, which constitutes a limitation for single-photon addition in bulk crystals. We call this particular wavelength of the pump the GVM wavelength, $\lambda_{GVM}$, satisfying  Eqs.~\eqref{GVM=PM}, at which in particular $\theta_{PM} = \theta_{GVM}$.


As shown in Fig.~\ref{fig:KDP-LN}, for the KDP crystal, the GVM and phasematching conditions are achieved for $\lambda_p = 415$ nm and  $\theta_{GVM} = \theta_{PM} = 67.74^{\circ}$,  while for LN, no pump wavelength satisfies Eqs.~\eqref{GVM=PM}. Table.~\ref{tab:Type-II-GVM} shows the different combinations of $\lambda_{GVM}$ and $\theta_{GVM}$ for four nonlinear crystals typically used in quantum optics experiments.

\begin{table}[h!]
    \centering
    \begin{tabular}{|c|c|c|}
\hline
 Crystal & $\lambda_{GVM}$ (nm) &  $\theta_{GVM}$ ($^{\circ}$)\\  
 \hline
 KDP & 415 & 67.74 \\ 
 \hline
  BBO & 585 & 30.96 \\
 \hline
 LN & - & - \\
 \hline
 BiBO & 647 & 24.12 \\
 \hline
  KTP & 711 & 46.84 \\
 \hline
    \end{tabular}
    \caption{Group velocity matching wavelengths $\lambda_{GVM}$ and angles $\theta_{GVM}$ for different nonlinear crystals in collinear degenerate Type-II PDC.}
    \label{tab:Type-II-GVM}
\end{table}

\begin{figure}[t!]
	\begin{center}
        \includegraphics[width=1.05\columnwidth]{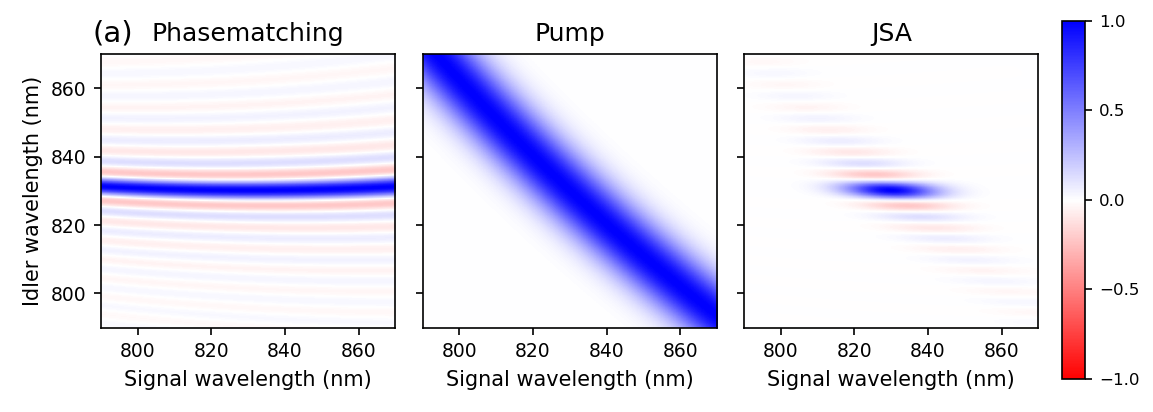}
		\includegraphics[width=0.95\columnwidth]{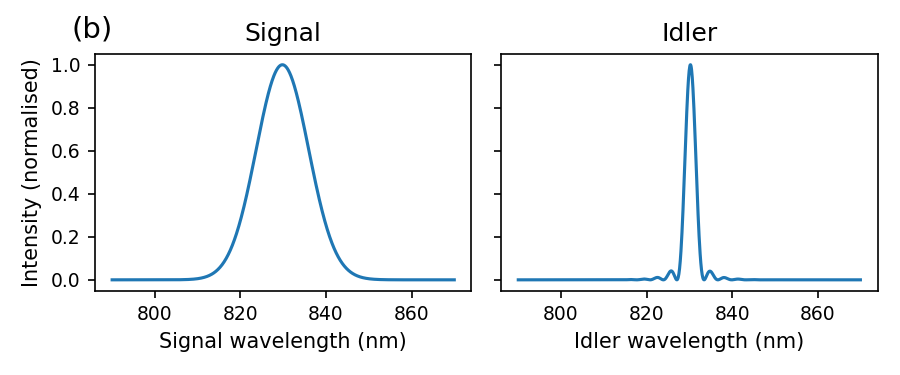}
	\end{center}
	\caption{\textbf{Collinear Type-II in KDP.} (a) JSA as the product of the phasematching and the gaussian pump. (b) First normalized signal and idler eigenmodes of the JSA.}
	\label{fig:JSA_KDP_T2_C_hg0}
\end{figure}



As seen in the previous section, the GVM condition is necessary but not sufficient for achieving $K=1$, as we should also have a phasematching bandwidth much smaller than the pump bandwidth. This can nevertheless be obtained by setting appropiately the crystal length, $L$ or the pump width, $\sigma_p$.


For the KDP crystal, the results are displayed on Fig.~\ref{fig:JSA_KDP_T2_C_hg0}, with a gaussian pump. The crystal length is set to $L=5$~mm, the pump bandwidth is $\sigma_p = 3$~nm, central wavelength of $\lambda_p = 415$~nm, and $\theta_{GVM}=67.74^{\circ}$ in this simulation.  

The singular value decomposition of the JSA is numerically performed, giving an effective number of modes $K=1.08$. For this set of parameters, the analytical expression~\eqref{eq:efficient_nb_mode_computed} obtained under gaussian approximations estimates $K=1.02$. The quantity $\sigma^2 \gamma L^2 (k_p' - k_i')^2 /2$ is computed to be around 10, which makes the condition on the pump and phasematching of  Eq.~\eqref{eq:condition_width} valid.


\begin{figure}[t!]
	\begin{center}
        \includegraphics[width=1.05\columnwidth]{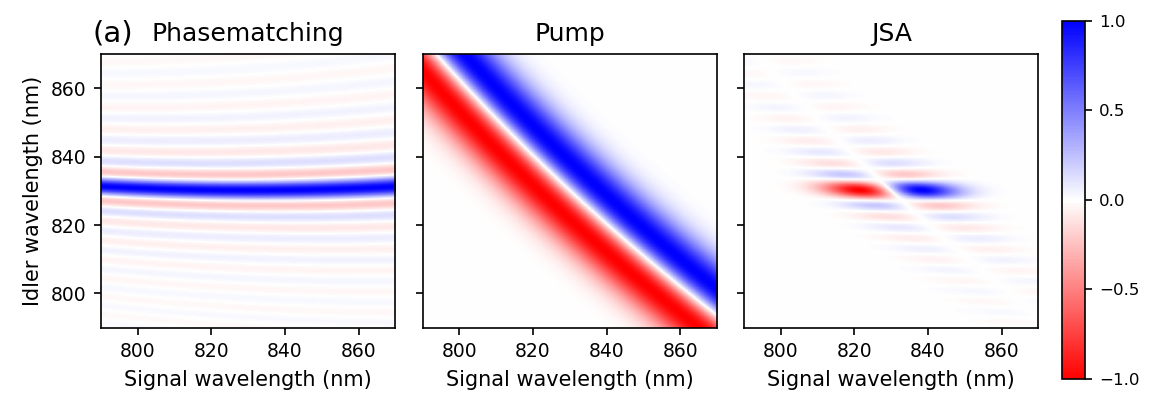}
		\includegraphics[width=0.95\columnwidth]{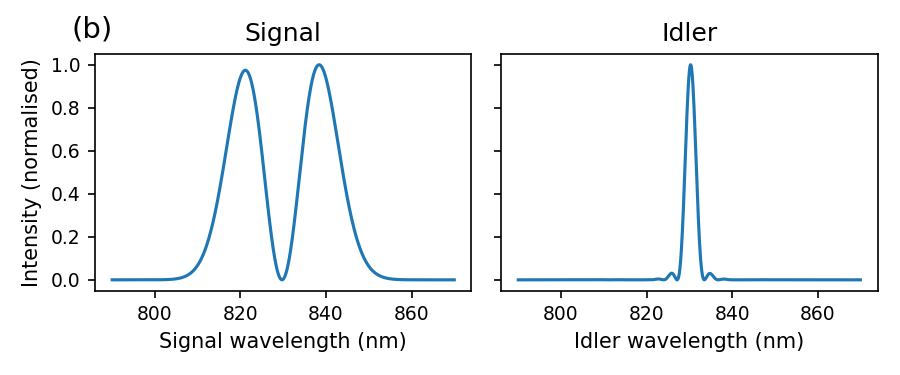}
	\end{center}
	\caption{\textbf{Collinear Type-II in KDP with first order pump.} (a) JSA as the product of the phasematching and the pump. The pump spectrum is a first order Hermite-Gaussian function. (b) First normalized signal and idler eigenmodes of the JSA.}\label{fig:JSA_KDP_T2_C_hg1}
\end{figure} 

We also compute the JSA function for a first order Hermite-Gaussian function as pump spectrum, cf. Fig.~\ref{fig:JSA_KDP_T2_C_hg1}~(a). As before, the first signal eigenmode has approximately the spectral shape as the pump's on Fig.~\ref{fig:JSA_KDP_T2_C_hg1}~(b). Hence, shaping the pump  allows for selecting the signal mode to which the photon is added. Here we obtain $K=1.17$, meaning that changing the pump spectrum can come at a cost on the effective number of modes of the process.

To sum up, the numerical simulations shows a  realistic configuration of mode-selective photon addition in a KDP crystal through collinear type-II PDC. Similar results are obtained for BBO, BiBO, and KTP crystals, in which the GVM condition of Eqs.~\eqref{GVM=PM} can also be satisfied.

\section{\label{sec:noncollinear} Single-mode photon addition in noncollinear PDC}

Single-mode photon addition can be also achieved in  noncollinear configurations. In this case, the noncollinear angle is a new degree of freedom that can be exploited to achieve the GVM condition at an arbitrary wavelength.

\subsection{\label{sec:noncollinear_typeII}Type-II}
 \begin{figure}[t!]
    \centering
    \includegraphics[width=1\columnwidth]{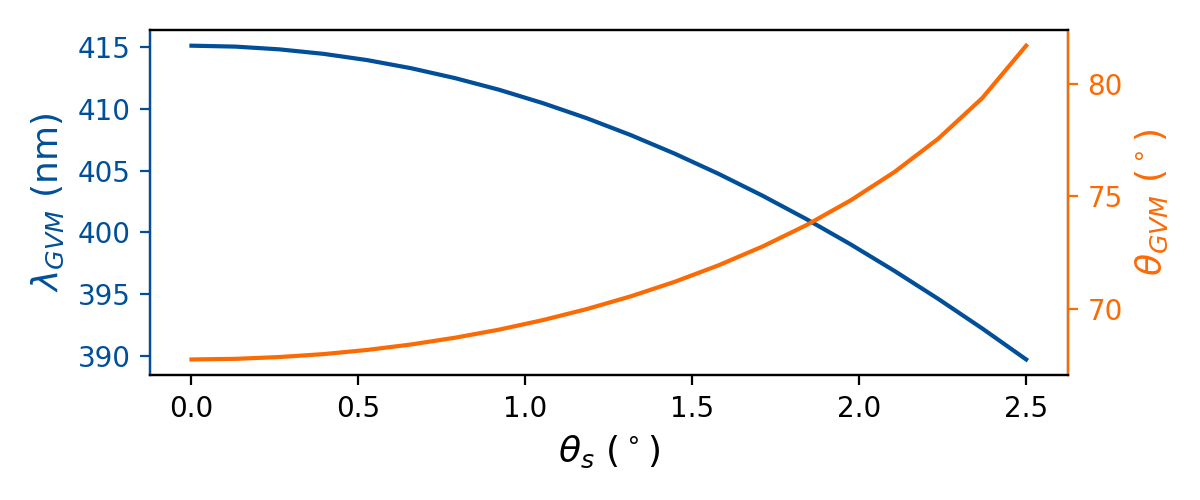}
	\caption{Group velocity matching wavelengths $\lambda_{GVM}$, and angles $\theta_{GVM}$, computed for different noncollinear angles $\theta_{s}$, for \textbf{KDP}  in  degenerate Type-II PDC.}\label{fig:KDP-GVM-T2-WL}
\end{figure}
 
The phase matching conditions for Type-II noncollinear PDC are \cite{10.1117/1.602464}: 
\begin{empheq}[left={\empheqlbrace\: }]{equation}
    \begin{aligned}\label{eq:T2-PM}
    2n_{e}(\lambda_{p}, \theta_{c}) &= n_o(\lambda_{s})  \cos(\theta_s) \\
    &\quad + n_{e}(\lambda_{i}, \theta_{c},\theta_{s}, \phi_s) \cos(\theta_i) \\
    n_o(\lambda_{s})  \sin(\theta_s) &= n_{e}(\lambda_{i}, \theta_{c},\theta_{s}, \phi_s) \sin(\theta_i) 
    \end{aligned}
\end{empheq}

where $n_{e}(\lambda_{i}, \theta_{PM},-\theta_{s}, \phi_s)$ is the refractive index of the idler field, $\theta_s$ (resp. $\theta_i$) is the angle of the signal (resp. idler) field with respect to the pump and $\theta_i = - \theta_s$. Equations in system~\eqref{eq:T2-PM} are solved simultaneously to find the phase matching angle $\theta_c = \theta_{PM}$. 

The index matching and group velocity matching curves intersect exactly at an unique pump wavelength for a given noncollinear angle $\theta_s$, as in the collinear case.
Fig.~\ref{fig:KDP-GVM-T2-WL} shows for each noncollinear angle $\theta_s$ the corresponding pump wavelength and group velocity matching angle for KDP. Similar results are obtained for BBO, BiBO and KTP.

In this noncollinear configuration, longitudinal and transverse components of the wavevector mismatches are given by:
\begin{empheq}[left={\empheqlbrace\: }]{equation}
    \begin{aligned}
        &\Delta k_{z} = k_p(\omega_p) - (k_s(\omega_s) + k_i(\omega_i)) \cos \theta_s \\
        &\Delta k_{\perp} = (k_i(\omega_i) - k_s(\omega_s) ) \sin \theta_s
    \end{aligned}
\end{empheq}

The first order Taylor expansion of the wavevector mismatch around the central frequencies gives:   
\begin{empheq}[left={\empheqlbrace\: }]{equation}
    \begin{aligned}\label{eq:T2NC-mismatch}
        \Delta k_z = \Delta k^{(0)}_z &+ (k'_p - k'_s \cos \theta_s) \Omega_s \\
        &+ (k'_p - k'_i \cos \theta_s) \Omega_i \\
        \Delta k_{\perp} = \Delta k^{(0)}_{\perp}  &-(k'_s \Omega_s - k'_i  \Omega_i) \sin \theta_s 
    \end{aligned}
\end{empheq}

where $k'_j$ 's are the inverse of the group velocities of the pump and downconverted fields evaluated at the central frequencies, and $\Omega_j = \omega_j - \omega_0$ with $j = s, i$, where $\omega_0$ is the signal and idler central frequencies. Here $\Delta k^{(0)}_z = k_p(2\omega_0) - \left(k_s(\omega_0) + k_i(\omega_0)\right) \cos \theta_s$ and $\Delta k^{(0)}_{\perp} = ( k_i(\omega_0)  - k_s(\omega_0) ) \sin \theta_s$ are the longitudinal and transverse components of the wavevector mismatch.

Here, both the transverse wavevector mismatch $\Delta k^{(0)}_{\perp}$ and the longitudinal wavevector mismatch $\Delta k^{(0)}_z$ should vanish for perfect phasematching. In Type-II PDC, the signal and idler fields have orthogonal polarizations, therefore their refractive indices are not equal. This makes it impossible to have both $\Delta k^{(0)}_{\perp}$ and $\Delta k^{(0)}_z$ equal to zero. Thus, only approximate phasematching can be achieved.

\begin{figure}[t!]
	\begin{center}
    \includegraphics[width=1.05\columnwidth]{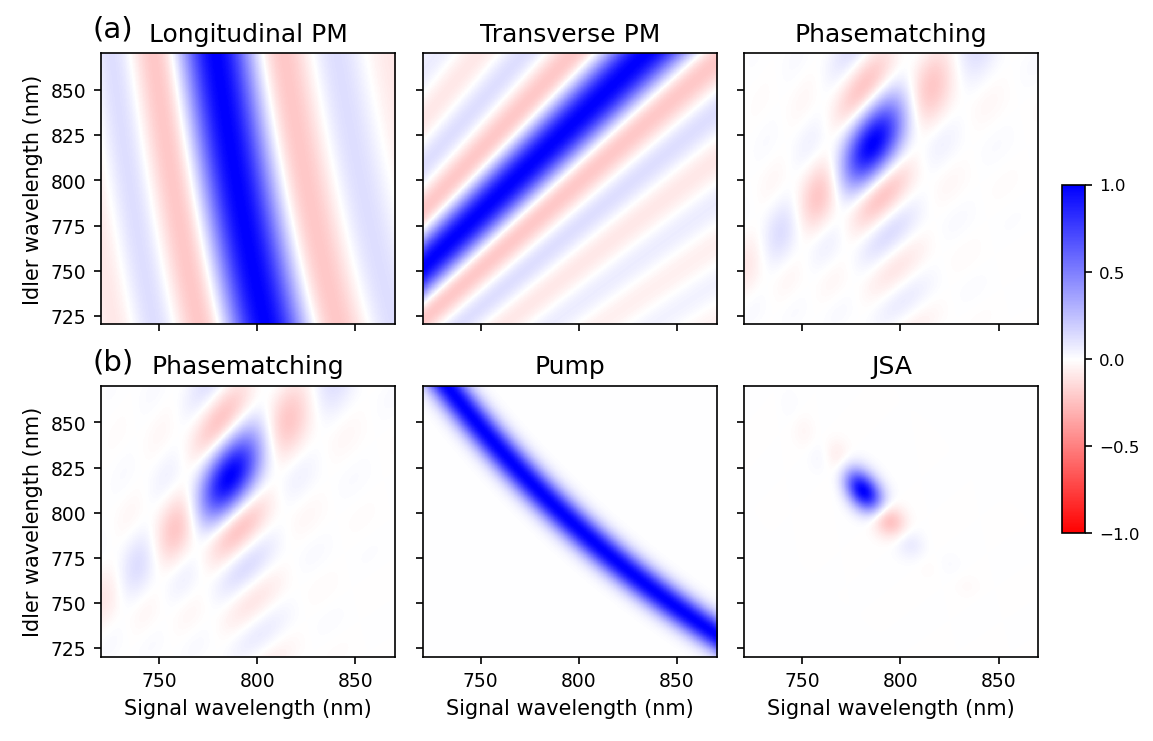}
    \includegraphics[width=1\columnwidth]{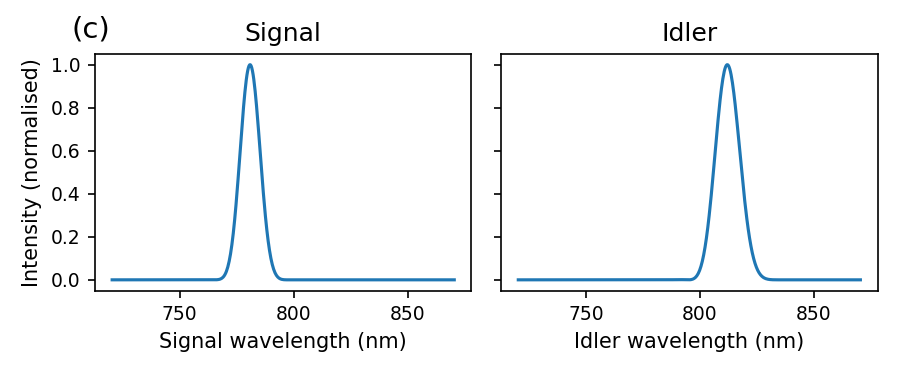}
	\end{center}
	\caption{\textbf{Type-II PDC in BBO.} (a) Phase matching as the product of the longitudinal phase matching and the transverse phase matching. (b) JSA as the product of the phase matching and the gaussian pump. (c) First normalized signal (left) and idler (right) eigenmodes of the JSA.}
	\label{fig:JSA_BBO_T2_NC}
\end{figure}

As given in  reference \cite{Uren2003}, the phase matching function $\phi(\Delta k_z,\Delta k_\perp)$ can be factorized into a product of its longitudinal and transverse parts:

\begin{equation} 
    \phi(\Delta k_z,\Delta k_\perp) \propto \phi_z(\Delta k_z) \phi^\perp(\Delta k^\perp)
\end{equation}

The longitudinal and transverse components of the phase matching functions can be approximately  written as:
\begin{empheq}[left={\empheqlbrace\: }]{equation}
    \begin{aligned}
        \phi_z(\Delta k_z) &= \exp \left(-\frac{\gamma \Delta k^2_z L^2}{4}\right) \\
        \phi^\perp(\Delta k^\perp) &= \exp\left(-\frac{\gamma (\Delta k_\perp)^2 \omega^2_0}{4}\right)
    \end{aligned}
\end{empheq}

In the frequency space ($\omega_s$,~$\omega_i$), the slope of the longitudinal phase matching function depends on the sum of signal and idler frequencies $\omega_s + \omega_i$ and its width depends on the length of the crystal $L$. Similarly, the slope of the transverse phase matching function depends on the frequency difference $\omega_s - \omega_i$ and its width depends on the beam waist, $w_0$. 
By changing the crystal length and the beam waist the overlap between the two can be engineered, in turn changing the width of the signal and idler fields. Therefore, after fixing the GVM wavelength and angle, the experimentally tunable parameters are the pump spectral width, the crystal length and the beam waist {$w_0$}. 

The results are shown in  Fig.~\ref{fig:JSA_BBO_T2_NC} for BBO  in the Type-II noncollinear configuration. The pump field was set as a gaussian with a spectral width of $\sigma_p = 5$~nm. For a  noncollinear angle of $\theta_s = 5.325^{\circ}$, we computed $\lambda_{GVM} = 398$~nm and $\theta_{GVM} = 49.1^{\circ}$. We have chosen a crystal length of $L = 0.3$~mm and a beam waist of $w_0 = 170\ \mu$m  for the simulation. From  Fig.~\ref{fig:JSA_BBO_T2_NC} it is clear that the transverse phasematching function is not centred around the desired central frequencies. As mentioned earlier in this section, it is due to the nonvanishing component of the transverse wavevector mismatch, as a result shifting the total phasematching function. We obtained the effective number of modes $K = 1.14$. The first signal and idler eigenmodes are displayed in  the bottom of Fig.~\ref{fig:JSA_BBO_T2_NC}.

\subsection{\label{sec:noncollinear_typeI}Type-I}

Finally, we treat the case of single-photon addition in degenerate Type-I noncollinear PDC. In this case, we can have perfect phasetmatching ($\Delta k^{(0)}_{\perp} = \Delta k^{(0)}_z=0$ in Eq.~\eqref{eq:T2NC-mismatch}). 

The phase matching condition for the degenerate Type-I PDC process is given by: 
\begin{equation}
    n_e(\omega_p, \theta_{c}) = n_o(\omega_s) \cos \theta_s
\end{equation}
where all the quantities involved have been defined above.

\begin{table}[h!]
    \centering
    \begin{tabular}{|c|c|c|}
\hline
 Crystal &  $\lambda_{GVM}$ (nm) & $\theta_{GVM}$ ($^{\circ}$) \\  
 \hline
 KDP & 517 & 41.15 \\ 
 \hline
 BBO & 771 & 19.83 \\
 \hline
 LN & 1012 & 44.95 \\
 \hline
 KTP & 919 & 24.98\\
 \hline
    \end{tabular}
    \caption{Group velocity matching wavelengths $\lambda_{GVM}$ and angles $\theta_{GVM}$ for different nonlinear crystals for degenerate Type-I PDC at $\theta_s=0^\circ$.}
    \label{tab:Type-I-GVM}
\end{table}

Table~\ref{tab:Type-I-GVM} shows  $\lambda_{GVM}$ and $\theta_{GVM}$ for different nonlinear crystals for degenerate Type-I PDC at $\theta_s=0^\circ$. 
\begin{figure}[t!]
    \centering
    \includegraphics[width=1\columnwidth]{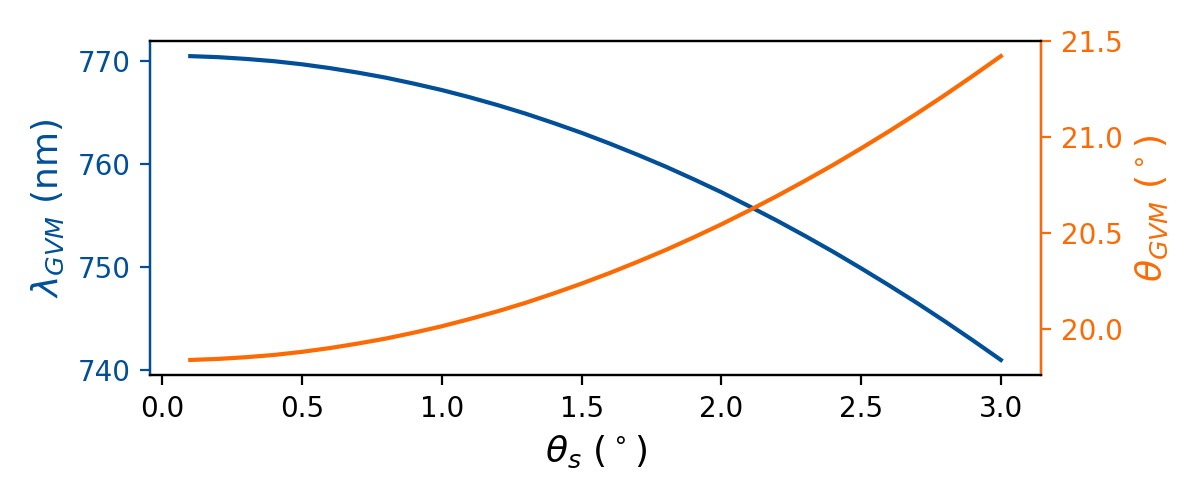}
	\caption{Group velocity matching wavelengths $\lambda_{GVM}$, and angles $\theta_{GVM}$, computed for different noncollinear angles $\theta_{s}$, for Type-I PDC in BBO}\label{fig:BBO-GVM-T1}
\end{figure}

In the case of BBO crystal, we show in  Fig.~\ref{fig:BBO-GVM-T1} the variation of $\lambda_{GVM}$ and $\theta_{GVM}$ with respect to noncollinear angles $\theta_s$. Besides, BiBO fulfills the conditions from $\theta_s=5^\circ$.


\begin{figure}[t!]
    \begin{center}
		\includegraphics[width=1.05\columnwidth]{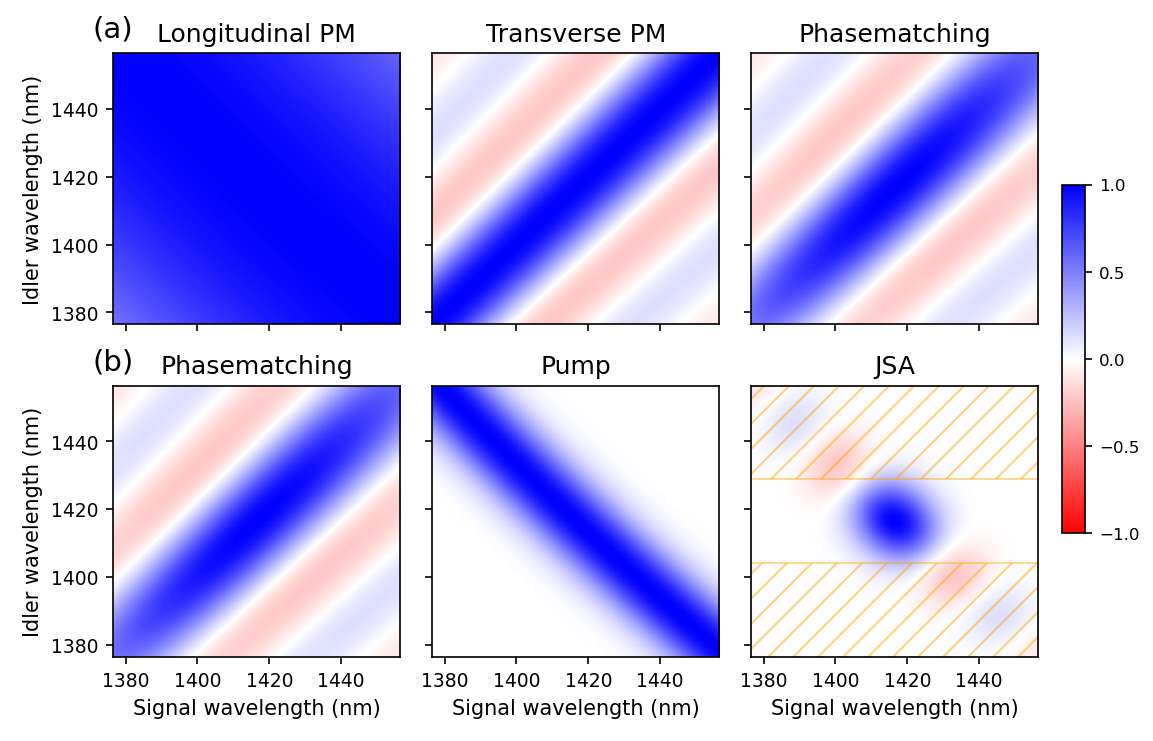}
		\includegraphics[width=1\columnwidth]{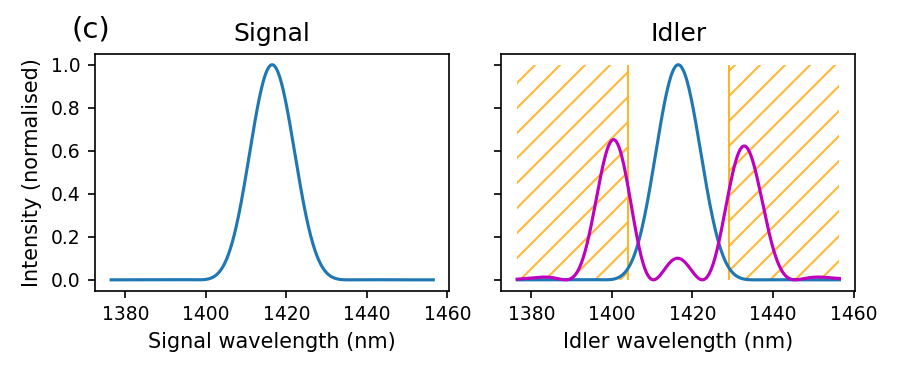}
	\end{center}
	\caption{\textbf{Type-I PDC in BiBO.} (a) Phase matching as the product of the longitudinal phase matching and the transverse phase matching. (b) JSA as the product of the phase matching and the gaussian pump. (c) In \textit{blue}: first normalized signal (left) and idler (right) eigenmodes of the JSA. In \textit{magenta}: the second idler eigenmode. The \textit{orange dashed area} represents the $25$ nm filtering on the idler field. For more details, see the main text.}
	\label{fig:BiBOType-IINC_JSA}
\end{figure}


In Type-I, $k'_s = k'_i$ and therefore Eq.~\eqref{eq:T2NC-mismatch} re-writes as (with $\Delta k^{(0)}_{\perp} = \Delta k^{(0)}_z=0$):
\begin{empheq}[left={\empheqlbrace\: }]{equation}
    \begin{aligned}
        &\Delta k_z = (k'_p - k'_s \cos \theta) (\Omega_s + \Omega_i) \\
        &\Delta k_{\perp} =  - k'_s \sin \theta (\Omega_s - \Omega_i)
    \end{aligned}
\end{empheq}
Numerical simulations are carried out for the uniaxial crystals KDP, BBO and LN as well as for biaxial crystals BiBO and KTP. 
Fig.~\ref{fig:BiBOType-IINC_JSA} shows the phasematching function and JSA for  BiBO  in the Type-I noncollinear configuration, the pump being a gaussian function with a spectral width of $\sigma_p = 6$ nm. Again for  $\theta_s = 5^{\circ}$, we computed  $\lambda_{GVM} = 708 $~nm and  $\theta_{GVM} = 8.02^{\circ}$. The central wavelengths are  $\lambda_s = \lambda_i = 1416 $~nm. The crystal length was set to $L = 1$ mm, and the beam waist to $w_0 = 550$ $\mu$m. The singular value decomposition for the JSA  is shown in Fig.~\ref{fig:BiBOType-IINC_JSA}.  The effective number of modes was calculated to be $K = 1.17$ under the sinc approximation.  The first signal and idler eigenmodes are displayed on Fig.~\ref{fig:BiBOType-IINC_JSA}.

For both Type-I and Type-II noncollinear PDC, mode-selectivity does not seem to be achievable from our studies. Indeed, increasing the order of the pump Hermite-Gauss mode by $1$ increases the effective number of modes by approximately $1$, resulting in non~single-mode photon addition.

\subsection{Filtering}

In order to obtain an effective number of modes $K$ closer to $1$, one may filter the idler field spectrally, so as to suitably select a unique signal mode.

For same set of input parameters as the Type-I BiBO computation, a $25$ nm wide spectral filter can be applied on the idler field. The effect of the filter translates into making the corresponding JSA part vanish, cf. Fig.~\ref{fig:BiBOType-IINC_JSA} (b). Also, on Fig.~\ref{fig:BiBOType-IINC_JSA} (c), most of the second idler eigenmode is filtered. This means that the probability of adding a photon to the first eigenmode is relatively increased, increasing the single-mode character of the addition process. Here, applying the filter yields an effective number of modes of $K = 1.03$, which is indeed a clear improvement. Besides, when $K$ is not too far from 1, most of the idler energy is contained in the first eigenmode, so that it does not decrease significantly the idler detection probability.

Similarly, adding a $5$ nm wide idler spectral filter to collinear Type-II PDC in KDP on both configurations Fig.~\ref{fig:JSA_KDP_T2_C_hg0} and Fig.~\ref{fig:JSA_KDP_T2_C_hg1} yields respectively $K=1.02$ and $K=1.05$ (with respect to $K=1.08$ and $K=1.17$ previously). One could obtain $K$ even closer to $1$ by choosing a thinner filter, but at the cost of lower photon counts. In the case of noncollinear Type-II PDC in BBO (cf. Fig.~\ref{fig:JSA_BBO_T2_NC}), we obtain $K=1.06$ (with respect to $K=1.14$ previously) with a $25$ mn wide filter.

Although filtering helps improving the single-mode character of the process, it does not clearly improve its mode-selectivity in the noncollinear configurations. Indeed, the success of filtering relies on the spectral distinguishability between the first idler eigenmode and the higher order ones. For noncollinear Type-I PDC in BiBO, with a first order Hermite-Gauss pump, $K=2$ and the first idler eigenmode highly overlaps with the second, which means filtering is not possible.

\section{Conclusion}

In this work, we developed a theoretical framework of the addition of a single photon to multimode light fields in order to generate non-Gaussian quantum states. We have investigated different PDC configurations that supports photon addition, with uniaxial and biaxial crystals (KDP, BBO, LN, BiBO, KTP).

For collinear Type-II PDC, mode-selective photon addition is shown to be achievable both analytically and numerically under group velocity matching and long enough crystal conditions. We prove that one can arbitrarily choose the unique mode in which the photon is added.

For noncollinear PDC, we extended the group velocity matching condition for both Type-I and Type-II processes, and show numerically that single-mode photon addition is possible. 

Moreover, filtering the idler field can be used to improve the single-mode character of photon addition.

Single photon addition is a promising operation to generate non-Gaussian multimode states of light. In particular, we anticipate that such state generation will be accessible in state-of-the art quantum optics experiments in the near-infrared and telecommunication wavelengths.

\section{Acknowledgements}
This   work   was   supported   by   the   European   Research Council under the Consolidator Grant COQCOoN (Grant No.  820079).


\bibliographystyle{apsrev4-2}
\bibliography{mybibliography}

\clearpage
\onecolumngrid
\appendix

\section{\label{app:A}Multimode photon addition}

\subsection{\label{app:A_cauchy}Proof that the Cauchy Schwarz inequality~\eqref{eq:cauchy_schwarz} cannot be saturated}

In section~\ref{multi_add_purity}, we obtained from Cauchy Schwarz inequality Eq.~\eqref{eq:cauchy_schwarz}:
\begin{equation}\label{eq:app_cauchy_schwarz}
|\bra{\phi} \hat{e}_1 \hat{e}_2^\dag \ket{\phi}|^2 \leq \bra{\phi} \hat{e}_1 \hat{e}^\dag_1 \ket{\phi} \bra{\phi} \hat{e}_2 \hat{e}^\dag_2 \ket{\phi} = (1+\bar{n}_1)(1+\bar{n}_2)
\end{equation}
This inequality is saturated iff $\hat{e}^\dag_1 \ket{\phi} \propto \hat{e}^\dag_2 \ket{\phi}$.\\
Let's write $\ket{\phi}$ over the Fock basis of the two addition eigenmodes\footnote{The other modes do not intervene in the computation.}:
\begin{equation}
\ket{\phi} = \sum_{n_1 \geq 0}\sum_{n_2 \geq 0} C_{n_1, n_2} \ket{n_1}\otimes\ket{n_2}
\end{equation}
where the complex coefficients $C_{n_1,n_2}$ ensure the normalisation.
\\So, the saturation condition $\hat{e}^\dag_1 \ket{\phi} \propto \hat{e}^\dag_2 \ket{\phi}$ re-writes:
\begin{empheq}[left={ \empheqlbrace}]{equation}\label{eq:recurrence}
\begin{aligned}
\: C_{n_1-1,n_2} \sqrt{n_1} &\propto C_{n_1,n_2-1} \sqrt{n_2} \qquad &\forall\, n_1 \geq 1, n_2 \geq 1 \\
\: C_{n_1-1,0} &= 0 \quad &\forall\, n_1 \geq 1 \\
\: C_{0,n_2-1} &= 0 \quad &\forall\, n_2 \geq 1 
\end{aligned}
\end{empheq} 
From this set of equations, it is easy to show recursively that:
\begin{empheq}[left={ \forall p \geq 0, \quad \empheqlbrace}]{equation}
\begin{aligned}
\: C_{n_1-1,p} &= 0 \quad &\forall\, n_1 \geq 1 \\
\: C_{p,n_2-1} &= 0 \quad &\forall\, n_2 \geq 1 
\end{aligned}
\end{empheq}  
This means that all coefficients $C_{n_1,n_2}$ must be zero, which is incompatible with the normalisation of $\ket{\phi}$. We conclude that Eq.~\eqref{eq:app_cauchy_schwarz} cannot be saturated.

\subsection{\label{app:A_genral_purity}Output state purity of multimode addition processes (general case)}
Following our developments in section~\ref{multi_add_purity}, let's show that the output is not pure for multimode addition processes (i.e. $K \neq 1$) generally, when one do not assume that only two eigenvalues are non-zero.
\\In this case, we have:
\begin{align}
\hat{\rho}^\mathrm{out}_s &= \sum_n \tilde{\lambda_n} \hat{e}^\dag_n \hat{\rho}^\mathrm{in}_s \, \hat{e}_n  \label{eq:app_output_density}
\\\mathrm{where}\quad \tilde{\lambda}_n &= \lambda_n/P \nonumber
\end{align}
The input $\rho = \ket{\phi}\bra{\phi}$ is still assumed pure. We find, using trace properties, that the output state purity writes:
\begin{equation}
\Tr [ (\hat{\rho}^\mathrm{out}_s)^2 ] = \sum_k \tilde{\lambda}_k^2 (1+\bar{n}_k)^2 + 2 \sum_{k>l}\tilde{\lambda}_k \tilde{\lambda}_l \left| \bra{\phi} \hat{e}_k \hat{e}_l^\dag \ket{\phi} \right|^2
\end{equation}
We apply the Cauchy-Schwarz inequality on the vector states  $\hat{e}^\dag_k \ket{\phi}$ and $\hat{e}^\dag_l \ket{\phi}$, as $\left| \bra{\phi} \hat{e}_k \hat{e}_l^\dag \ket{\phi} \right|^2 \leq (1+\bar{n}_k)(1+\bar{n}_l)$.\\We obtain:
\begin{equation}
\label{eq:app_cauchy_swartz}
\Tr [ (\hat{\rho}^\mathrm{out}_s)^2 ] \leq \sum_k \tilde{\lambda}_k^2 (1+\bar{n}_k)^2 + 2 \sum_{k>l}\tilde{\lambda}_k \tilde{\lambda}_l (1+\bar{n}_k)(1+\bar{n}_l) = \left( \sum_k \tilde{\lambda}_k (1+\bar{n}_k)\right)^2 = 1
\end{equation}
where we used the fact that taking the trace of Eq.~\eqref{eq:app_output_density} yields 1.
\\Again, looking at the saturation of the Cauchy-Schwarz inequality leads to a similar set of equations as in Eq.~\eqref{eq:recurrence} for k and l fixed (except that the state $\ket{\phi}$ is decomposed over the full Fock space). Solving the recurrence equations for a given k, l, shows that the saturation condition can't be satisfied. Thus, the purity of the output density matrix is strictly lower than $1$, meaning that the output state is not pure.

\section{\label{app:B_derivation_K}Proof of the derivation of the analytical from of the effective number of modes $K$}

In this section, we show the analytical formulae  Eq.\eqref{eq:efficient_nb_mode_computed}.\\
The definition of the effective number of modes $K$ is recalled:
\begin{equation}\label{eq:efficient_number_of_modes_app}
K = \dfrac{(\sum_n \lambda_n)^2}{\sum_n \lambda_n^2}
\end{equation}
The JSA function is conveniently expressed in the gaussian form:
\begin{equation}\label{app:JSA_gaussian}
R(\omega_s, \omega_i) = D \exp\left[-\frac{1}{2}x^T V x \right]
\end{equation}
where $x^T = (\omega_s, \omega_i)$, $V$ is a 2x2 matrix, and $D$ is a proportionality coefficient.
The Schmidt decomposition of the JSA function into signal and idler frequency eignemodes is expressed as:
\begin{equation}\label{app:JSA_function_Schmidt_decomposition}
R(\omega_s, \omega_i) = \sum_{n \geq 1} \sqrt{\lambda_n} \psi_n^*(\omega_i) \varphi_n(\omega_s)
\end{equation}
Using Eqs.~\eqref{app:JSA_gaussian} and~\eqref{app:JSA_function_Schmidt_decomposition}, we obtain two expressions of the integral of $R$:
\begin{align}
\sum_n \lambda_n = \int d\omega_s d\omega_i |R(\omega_s, \omega_i)|^2 = D^2 \frac{2 \pi}{\sqrt{\det(2V)}} \label{eq:sum_lambda}
\end{align}
where we used the orthonormal properties of the eigenmodes for the left most member, and we performed the integrals using the following general expression for Gaussian integrals to get the right most member:
\begin{equation}
\int \dots \int \exp\left[-\frac{1}{2}q^T M q \right] dq_1 \dots dq_n = \frac{(2 \pi)^{n/2}}{\sqrt{\det(M)}}
\end{equation}
Note that we assumed that $V$ is real, which can be checked on Eq.~\eqref{eq:JSA_gaussian_form}. 
\\Similarly, we find:
\begin{equation}
\sum_n \lambda_n^2 = \int d\omega_s d\omega_s' |A(\omega_s, \omega_s')|^2 = D^4 \frac{(2 \pi)^{2}}{\sqrt{\det(W)}}\label{eq:sum_lambda_squared}
\end{equation}
where $W$ is defined as a 4x4 matrix such that:
\begin{equation}\label{eq:definition_of_W}
R(\omega_s, \omega_i) R^*(\omega_s', \omega_i) R^*(\omega_s, \omega_i') R(\omega_s', \omega_i') = D^4 \exp\left[-\frac{1}{2}X^T W X \right]
\end{equation}
Now, substituting equations \eqref{eq:sum_lambda} and \eqref{eq:sum_lambda_squared} into the definition \eqref{eq:efficient_number_of_modes_app}, we obtain the expression of the effective number of modes through the matrices defined above:
\begin{equation}\label{eq:efficient_number_of_modes_matrices}
K = \frac{\sqrt{\det(W)}}{4\det(V)}
\end{equation}
Let us now specify the expression of $K$ to this problem. $V$ can be deduced from the JSA expression~\eqref{eq:JSA_gaussian_form}:
\begin{align}
V &= \frac{1}{\sigma^2} \begin{pmatrix}
1 + r_s^2 & 1 + r_s r_i \\
1 + r_s r_i & 1 + r_i^2 
\end{pmatrix} \label{eq:matrix_V}\\
\mathrm{with}\quad r_j &= \sigma L \sqrt{\frac{\gamma}{2}} \, | k_p' - k_j'|,\quad \mathrm{for} \ j = i, s \nonumber
\end{align}
where the $r_j$ coefficients are the adimensioned parameters of the problem. The definition~\eqref{eq:definition_of_W} of $W$ leads to:
\begin{equation}\label{eq:matrix_W}
W = \frac{1}{\sigma^2}\begin{pmatrix}
2(1+r_s^2)&1 + r_s r_i&0&1 + r_s r_i\\
1 + r_s r_i&2(1+r_i^2)&1 + r_s r_i&0\\
0&1 + r_s r_i&2(1+r_s^2)&1 + r_s r_i\\
1 + r_s r_i&0&1 + r_s r_i&2(1+r_i^2)\\
\end{pmatrix}
\end{equation}
Computing the determinant of matrices~\eqref{eq:matrix_V}~and~\eqref{eq:matrix_W}, and substituting them into equation~\eqref{eq:efficient_number_of_modes_matrices}, we end up with an analytical expression for $K$ under gaussian approximations:
\begin{align}
K &= \sqrt{\frac{(1+r_s^2)(1+r_i^2)}{(r_s-r_i)^2}} \label{eq:efficient_nb_mode_computed_app} \\
\mathrm{with}\quad r_j &= \sigma L \sqrt{\frac{\gamma}{2}} \, | k_p' - k_j'|,\quad \mathrm{for} \ j = i, s \nonumber
\end{align}
\end{document}